# A proposed signal discovery method in interstellar communication

*William J. Crilly Jr.*

Green Bank Observatory, West Virginia, USA

*Abstract*—Experiments conducted since 2018, using three geographically spaced synchronized radio telescopes, and a radio interferometer, indicate the presence of anomalous narrow bandwidth pulse pairs, conjectured to be sourced from a celestial direction near the star Rigel. Many explanatory hypotheses are possible. In the current work, a measurement method is proposed and implemented to attempt to provide high levels of statistical power of narrowband pulse pair observations, while minimizing interferometer instrument adjustments that might bias the experiment. Using the proposed method, twenty pulse pairs having statistical power at 5 to 10 standard deviations, and one pulse pair at 22 standard deviations of mean shift of pulse pair count, were observed in the prior celestial direction range, while using other celestial directions as an experimental comparison group. High levels of standard deviations of pulse pair count mean shift, observed in this 92 day experiment, imply the falsification of a Gaussian noise hypothesis. Alternate and auxiliary hypotheses, and further experiments, are sought to try to explain the narrow bandwidth pulsed signals.

## I. Introduction

The identification of intentionally discoverable communication signals, without prior coordination, inherently involves a tradeoff between channel information rate and ease of discovery. Signals having high sustained information rate, at the Shannon energy and capacity limits, are practically indistinguishable from Additive White Gaussian Noise (AWGN).[1] Assuming that discoverable high information rate interstellar communication exists, transmitter modulation systems are speculated to be designed to have various characteristics, e.g. energy-efficient and readily discoverable pulse position and frequency position modulation, designed to propagate through an Interstellar Coherence Hole (ICH) [2][3][4].

Discoverability is expected to be enhanced by the transmitting entity repeating transmitter parameters, e.g. pulse duration and pulse occupied bandwidth, and by transmitting simultaneous narrow bandwidth pulses on closely spaced RF frequencies.

An objective of the work in this project is to apply communication principles to propose and implement methods that might help indicate the presence of intentionally discoverable high information rate celestial communication signals.

**Prior work**

During experiments conducted with four radio telescopes, starting in 2017, anomalous narrow bandwidth $\Delta t$ $\Delta f$ polarized pulse pairs have been repeatedly observed, beginning in 2018, having an apparent celestial source location at 5.25 ± 0.15 hr Right Ascension (RA) and -7.6º ± 1º Declination (DEC). Prior AWGN explanatory hypotheses have been repeatedly falsified. [5][6][7][8][9][10] Radio frequency interference (RFI) is ameliorated using an automated RFI discovery, excision and reporting algorithm, improved in the current work.[6][9]

## II. Hypothesis

**Hypothesis**

Interstellar communication signals comprising narrow bandwidth $\Delta t=0$ $\Delta f$ polarized pulse pairs, observed within a celestial coordinate range of 5.25 ± 0.15 hr RA and -7.6º ± 4º DEC, are expected to indicate measurements having statistical power indistinguishable from measurements of AWGN, during long duration algorithm-based RFI-excised experiments.

**Falsification**

Statistical power that implies falsification of an AWGN explanatory hypothesis, compels the development of alternate and auxiliary hypotheses and their associated experimental methods.[11]

**Comparison group**

The comparison group in this experiment comprise pulse pair counts in pointing directions outside the prior celestial direction, expected to be due to artifacts of random sky and measurement noise, statistical variation, natural object emissions, RFI leakage, and unknown causes.

## III. Method of Measurement

**Objectives**

1. Minimize experimenter adjustments of instruments during observation data acquisition and post-processing.
2. Emphasize the importance of large values of statistical power of the measurements of anomalies, given the hypothesized signal characteristics, underlying known physical processes, and experimental comparison group.
3. Minimize RFI false positives using a machine-only RFI detection, excision and reporting system.
4. Simplify the experimental system to facilitate experimental replication, troubleshooting and search for false positives.
5. Perform long duration tests, to reduce diurnal effects and RFI, and to seek repetition of celestial direction anomalies.
6. Improve the precision of RA measurements.

**Measurements**

**Right Ascension (RA)**

The RA of a recorded pulse is calculated at the Modified Julian Date (MJD) that the perpendicular bisecting plane of the East-West interferometer baseline intersects a measured 180.0º azimuth. The baseline perpendicular azimuth value was measured using solar shadow measurements, and assumes identical antenna reflector surface feed illumination. The interferometer has a 0.117 hr RA aliasing period due to integer-wavelength ambiguity of geometric space delay. [12] RA measured values are retained with full resolution, and quantized in 0.015 RA hr size bins. The RA is binned to a

---

William J. (Skip) Crilly Jr. is a Volunteer Science Ambassador in Education & Public Outreach of the Green Bank Observatory. email: wcrilly@nrao.edu

resolution that provides approximately eight RA bins per integer-wavelength space delay aliasing period. Antenna element full width half maximum (FWHM) beam width spans approximately four wavelength-aliased pointing directions. Aliasing is expected to increase pulse detection rate due to a transmitter transiting integer-wavelength aliased RA directions, while within the FWHM antenna response.

### $\Delta_{EW}\varphi$

The difference in signal phase between West and East antenna signals, is measured in each 3.7 Hz wide FFT bin, at in-phase and quadrature-phase (IQ) baseband, during a 0.27 s duration. West and East pulse recorded signals each exceed a Signal to Noise Ratio (SNR) threshold. The $\Delta_{EW}\varphi$ phase measurement is used to seek repetitive nearly equal angles of arrival over a long duration. Near-zero $\Delta_{EW}\varphi$ values indicate angle of arrival near-perpendicular to the East-West baseline, or in directions aliased by integer-wavelength differential space propagation distance to antenna elements, and by RF frequency aliasing due to -82 $\tau_{INT}$ instrument delay, or in combination.

The $\Delta_{EW}\varphi$ measurements contain several phase noise contributions, after assigning a $\Delta_{EW}\varphi$ measurement to an RA bin, based on the MJD at the time of pulse reception. These phase noise contributions are described as follows:
1. RF frequency-dependent phase shift is caused by RA values that are not centered in a ±0.0075 hr RA bin, at which noise-free, zero valued phase indicate a 180.0 azimuth direction of arrival. This phase noise is calculated to be distributed within ± 0.40 radians at 1425 MHz, i.e. $2\pi$ * 0.0075 hr divided by the 0.117 hr one wavelength geometric delay aliasing period.
2. RF frequency-dependent phase shift is caused by instrument differential delay from each antenna element to the phase detector, and variable due to various instrument imperfections, e.g. cable delay variation, transition reflections. This instrument delay induced phase noise is expected to be distributed over [-$\pi$:$\pi$]. Near-zero phase values occur at near-periodic interval RF frequencies, due to aliasing.
3. The noise vector, relative to the signal vector, affects the measured $\Delta_{EW}\varphi$. At a signal power to noise power ratio of 10, phase noise is estimated to be distributed in the range of $\pm\tan^{-1} 10^{-0.5}$ = ±0.3 radians.
4. Phase measurement is offset due to phase detector measurement uncertainties, e.g. IQ channel cross-coupling, local oscillator quadrature phase offset, typically within ±0.1 radians.

Each phase noise category is expected to have an associated probability density function. It is expected that if phase noise causes 1, 2, and 3 have independent, zero mean distributions, then the cascaded convolution of the density functions may be approximated with a zero mean Gaussian distribution.

Phase noise cause 4 is expected to generally result in an offset from zero phase measurement, that may be time-varying.

The overall effect of phase noise on $\Delta_{EW}\varphi$ is expected to provide an apparent concentration of signal phase measurements, in an RA bin, at an offset from zero measured phase. The analysis of phase noise contribution is a topic of further work.

### $\Delta_{\Delta f}\Delta_{EW}\varphi$

The difference in $\Delta_{EW}\varphi$ phase values of simultaneous pulses in a pair, separated in RF frequency by $\Delta f$, is used in a post-processing filter to seek pulse pairs having repetitive near-equal angles of arrival. This measurement is corrected for interferometer instrument delay as follows:

$$\Delta_{\Delta f}\Delta_{EW}\varphi = \Delta_{\Delta f}\Delta_{EW}\varphi_{MEASURED} + 2\pi\,\Delta f\,\tau_{INT}, \quad (1)$$

in radians, where $\Delta f$ is the frequency separation between $\Delta t=0$, $\Delta f$ pulses in a pulse pair, in MHz, and $\tau_{INT}$ is the measured interferometer instrument delay in microseconds. The $\Delta_{\Delta f}\Delta_{EW}\varphi$ measurement has phase noise expected to primarily be due to phase noise contributions 3 and 4 described above, as 1 and 2 are expected to have low values, due to the $\Delta_{\Delta f}$ calculation.

### Cohen's d of pulse pair count

The statistical power in this work is provided by the measurement of the normalized mean shift of a chosen RA population event count, relative to other RA populations. **[13]** Populations are contained in 1,600 RA bins in 24 hr RA. The mean count shift is normalized by the calculated standard deviation of the tested population. In this work, a binomial distribution is used to reflect the AWGN-caused presence of pulses in an RA bin. RA bin binomial distribution event probabilities are calculated based on the duration that an RA bin is sampled in the post processed files. The number of RA bins is chosen to separately resolve multiple pointing directions within the RA aliasing period of 0.117 hr RA.

### $Log_{10}$ Likelihood of composite SNR (LLSNR)

The LLSNR measurement calculates the likelihood of observing SNR values that exceed 8.5 dB SNR, assuming a Rayleigh amplitude distribution in Gaussian noise, using

$$LLSNR = 6.149 - 0.4343(10^{SNR\_East/10} + 10^{SNR\_West/10}), \quad (2)$$

where SNR_East and SNR_West are the element SNR measurements of a pulse, in dB. Pulse pairs sum four constituent SNR values.

### Signal discovery method

The method to seek discovery of transmitted $\Delta t=0$ $\Delta f$ polarized pulse pairs comprises the following five steps and their associated presentations of results in this work.
1. Measure $\Delta_{\Delta f}\Delta_{EW}\varphi$ of 3.7 Hz bandwidth 0.27 s integrated pulse energy, having high SNR thresholds, in long duration tests using a constant interferometer baseline. Sort pulses by increasing value of $|\Delta_{\Delta f}\Delta_{EW}\varphi|$. Low values of $|\Delta_{\Delta f}\Delta_{EW}\varphi|$ are expected from pulse pairs having simultaneous $\Delta f$ spaced components with close to the same angle of arrival. Calculate statistical power. **Fig. 1**
2. Examine near-zero indicated $\Delta_{EW}\varphi$ values, within the dataset sorted by increasing $|\Delta_{\Delta f}\Delta_{EW}\varphi|$ in **step 1**. Seek a cluster, comprising excess common-phase pulses, expected to be ±n2$\pi$ aliased due to MJD-dependent multi-$\lambda$ differential propagation distance, and ±m2$\pi$ aliasing due to frequency dependent phase shift caused by -82 ns differential instrument delay between antenna elements. A near-zero valued $\Delta_{EW}\varphi$ cluster of pulse pairs observed in an RA bin is expected to indicate an anomalous received pulse pair rate, relative to comparison group RA bin pulse pair rates, if anomalous pulses are present. **Fig. 2**
3. Measure the statistical power of excess pulse pair count, while sorting by increasing $|\Delta_{EW}\varphi|$. **Fig 3**

4. Measure the statistical power as in **step 3**, after offsetting the $\Delta_{EW}\phi$ measurement by the near-zero offset value of a candidate $\Delta_{EW}\phi$ cluster observed in **step 2**. **Fig 4**
5. Examine associated measurements of the candidate pulse pairs across RA, e.g. MJD, pulse pair **Δf**, RF frequency, to develop auxiliary and alternate hypotheses that might explain observed phenomena. **Figs. 5-12**

## Measurement system

### Antennas

Two ten foot diameter offset-fed parabolic reflectors are spaced 33.0 wavelengths, at 1425 MHz, along an East-West baseline in a heavily forested rural area. Crossed folded dipoles are located in a 1.2 foot diameter conical reflector feed. Right-hand circular polarized signal energy is chosen due to propagation reasoning, [14] and is produced using a quadrature hybrid phase shift component, after low-noise amplification at each feed. RF cables to indoor receivers are different in length, to provide the correlator with two signals that ameliorate the zero time offset common-mode instrument noise often present. RA periodicity is measured at a period of 0.117 RA hours, using measured 33.0 $\lambda$ element spacing, and at a sun shadow measured 180.0º baseline bisecting plane azimuth. -82 ns differential instrument delay was measured using a test antenna at each reflector surface, connected to a vector network analyzer. Antenna beam overlap uses NRAO 5690 as a celestial calibration object. [15]

### Narrow bandwidth receivers

Signals are down-converted to baseband using a 1425 MHz local oscillator disciplined to GPS satellite signals. In-phase quadrature (IQ) baseband signals comprising 50 MHz RF bandwidth of each antenna element are digitized in four synchronously triggered and sampled analog to digital converters. Signals are sampled at 62.5 Megasamples per second, at eight bit resolution, triggered by a GPS-disciplined pulse having a three second interval. Fast Fourier Transforms are calculated using the FFTW3 library [16] to produce 3.7 Hz bandwidth channels providing complex amplitude. SNR is calculated using a 954 Hz bandwidth noise measurement surrounding the SNR-calculated 3.7 Hz FFT bin. Details are described in **IV. C. *Receivers*** in [5].

Narrowband signal phase per element is calculated using $\tan^{-1}(Im/Re)$, where Im and Re are the imaginary and real components of the complex amplitude measured in each FFT bin.[2] Narrowband signal phase is used to facilitate a measurement associated with pulse angle of arrival at the interferometer baseline.

### Cross-correlator

A 50 MHz bandwidth, 0.27 s integration FX cross-correlator [12] operates in parallel with the narrow bandwidth signal processing. The cross-correlation is used to observe celestial natural objects, and provides means to evaluate receiver system performance and calibration.

### Pulse pair record storage

Measurements are thresholded at 8.5 dB SNR of each antenna pulse and stored in files having four hour MJD duration. The MJD measurement is sourced from a GPS receiver, providing a low frequency digital code to one channel of the four channel digitizer, to ensure that pulse records have properly synchronized MJD. Stored measurement files are transferred over a wired network to a post-processing machine. Receiver data storage operates continuously and is battery backed up.

### Post-processing

Post-processing provides for the setting of filter threshold parameters, statistical analysis and presentation of results. Post-processing filters include SNR threshold, Δf range, $\Delta_{\Delta f} \Delta_{EW}\phi$ range, $\Delta_{EW}\phi$ range, RA bin count and RFI margin. SNR filtering is a natural setting in receivers to reduce noise induced measurement results, and is set in this work based on observed SNR measurements of candidate pulses. RA bin resolution at 0.015 hr RA provides approximately eight RA bins across the calculated period of RA aliasing. Post-processing software includes the generation of the image files of the figures in this work.

### RFI amelioration

The 50 MHz instantaneous spectrum is separated into spectral segments each having 256 FFT bins, comprising 954 Hz. Each spectral segment keeps a count of potential RFI pulses, during the four hour duration of the file. Details of the RFI system are described in **Measurement System *RFI Excision*** in [9]. An improvement in the RFI amelioration in the current work has been to add a look-forward algorithm, to excise suspected RFI segments that are identified and recorded at a later time in the four hour duration RFI file.

## IV. OBSERVATIONS

Observations reported here are based on the O7a and O7b observation runs in this work, having a dataset comprising 92 days, during 105 days. The 13 day difference is due to a time period that the system was powered down. For each day that the system was powered up, data files were produced and post-processed. The RA range between 1.5 to 9.0 hours was examined in post-processing, using data from the narrow bandwidth channelized receivers, complex cross-correlator, and metrology.

Implementation of the first four steps in the proposed signal discovery method are presented in **Figs 1-4**. **Figs 5-12** display associated measurements that might contribute to understanding the possible causes of the narrow bandwidth pulse pair phenomena.

Salient aspects of measurements are described in the text below each figure. Additional notes follow.

**Fig. 1: Step 1** of the discovery method is important to seek RA directions of interest, with a minimal number of observer adjustments.

Common mode residual broadband noise coupling at the -5x16 ns = -80 ns correlator delay tap causes fluctuations, over RA, in visibility amplitude, and non-linear phase. Pulsed 3.7 Hz bandwidth phase and amplitude measurements are not expected to be noticeably affected by residual correlator common mode near-zero-tap broadband noise, due to the high 50 MHz to 3.7 Hz bandwidth factor.

**Fig. 2: Step 2** of the proposed presentation shows an apparent cluster of data points near $\Delta_{EW}\phi \approx 0.07$ radians. This value is in the range that might be due to phase detector measurement offset. Further work is planned to investigate this possibility.

**Fig. 3: Step 3** has the primary purpose of examining the statistical power of normalized mean shift of pulse pair count, without introducing an offset in the $\Delta_{EW}\phi$ measurement, as the offset is an observer setting which might bias the statistical power calculations.

**Fig. 4: Step 4** is used to determine the statistical power of $\Delta_{EW}\phi$ observations in an RA direction of interest, using the apparent phase detector offset observed in **Fig. 2**. The Cohen's **d** is expected to increase significantly in **Fig. 4** compared to **Fig. 2**, because the cluster of points in **Fig. 2** is highly unlikely to be caused by AWGN.

**Figs. 5-12 Step 5** describes associated measurements that might help in the development of alternate and auxiliary hypotheses to explain the phenomena.

A proposed signal discovery method in interstellar communication

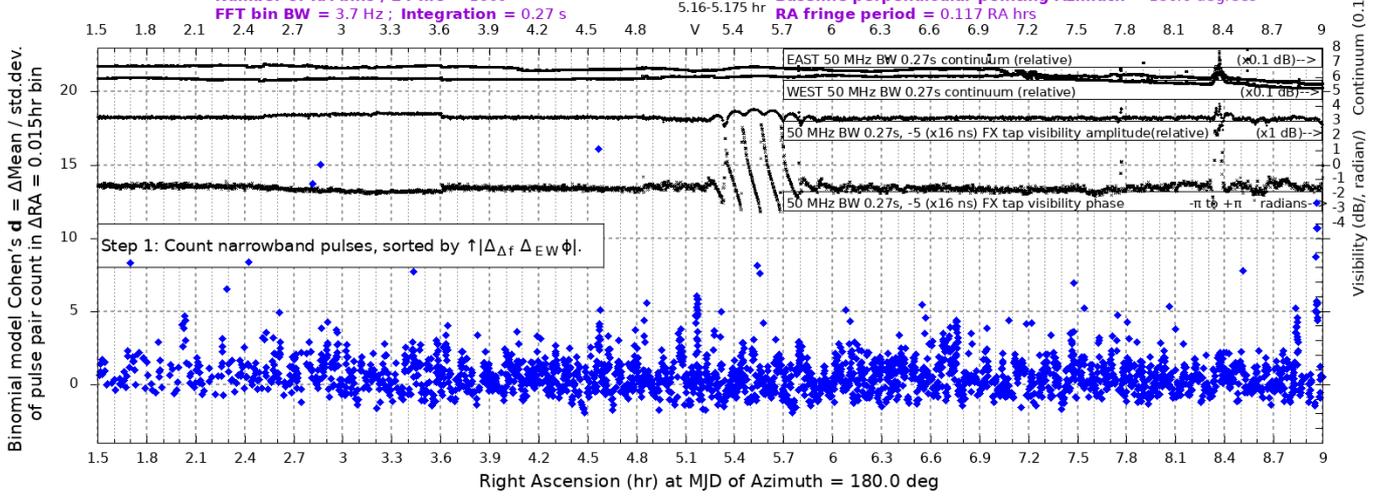

**Fig. 1: Step 1:** Increasing $|\Delta_{\Delta f} \Delta_{EW}\varphi|$ sorted counts of the $\Delta f$-spaced and East-West phase pulses are plotted vs. RA, at a fixed 180.0º telescope azimuth. The pulse pairs in each RA bin are counted on a data set sorted by increasing absolute value of difference in $\Delta_{EW}\varphi$ values of the -82 ns instrument phase delay-corrected $\Delta f$ pulses in each pulse pair. Statistics are calculated based on an expected binomial distribution, RA binned, and modeled as AWGN, seeking a repetitive common angle of arrival. 50 MHz bandwidth continuum power, and complex visibility are plotted in the four upper traces. Periodic RA visibility phase response is indicated in the observation of the emission nebula Orion A, **[15][17]** at 5.5 hr RA.

Among the 50 highest Cohen's **d** values in the 3,417 point data set, i.e. those exceeding 4.4 Cohen's **d** standard deviation, 12 values are in the 5.160-5.175 hr RA bin, 10 values are in the 8.955-8.97 hr RA bin, and 28 values are distributed among RA bins with no more than two values in a bin. 473 RA bins of the 500 bins in the 1.5 to 9.0 hr RA range do not have Cohen's **d** values higher than 4.4 standard deviations. The Cohen's **d** of the 5.16-5.175 hr RA bin appears to indicate high statistical power, using the method in **Step 1**.

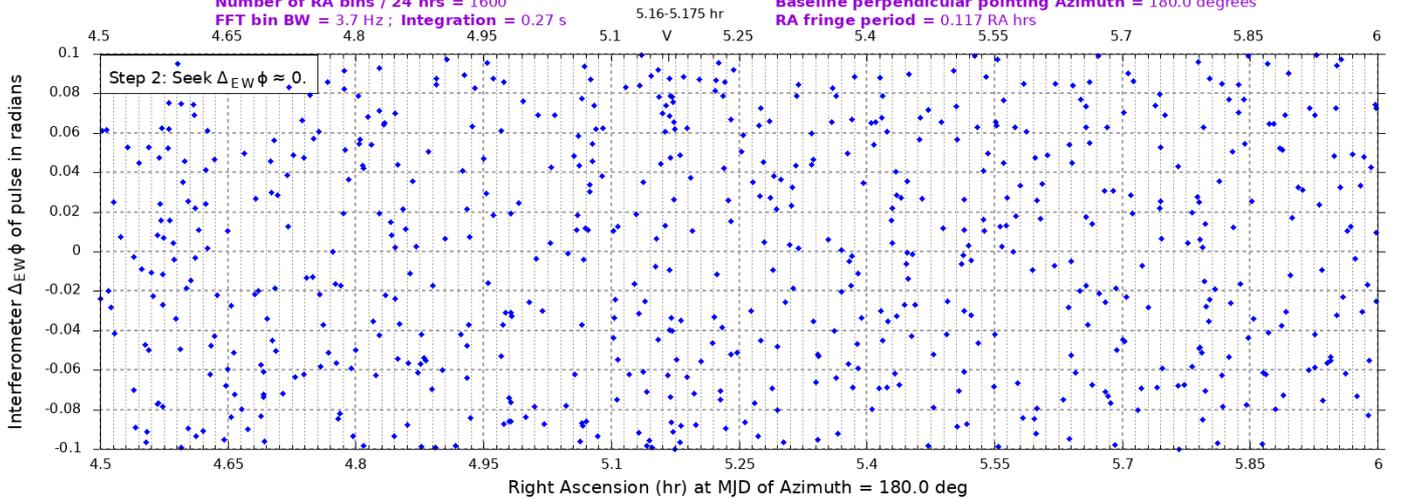

**Figure 2: Step 2:** A residual zero offset of the $\Delta_{EW}\varphi$ measurement of constituent pulses in aliased pulse pairs is affected by residual instrument uncertainties, e.g. phase detector zero offset, IQ imbalance. The narrowband pulse phase values were examined to seek a concentration of pulse pairs near zero radians $\Delta_{EW}\varphi$. Near zero $\Delta_{EW}\varphi$ values are expected to reflect repetitive common directions of arrival of constituent pulses. A cluster of measurements surrounding 0.07 radians was observed in the 5.16 to 5.175 hr RA bin. This zero offset value was used in the increasing $|\Delta_{EW}\varphi+\text{offset}|$ sorted Cohen's **d** measurements in **Fig. 4**. In **Fig. 3**, an offset of zero radians is applied, to count increasing $|\Delta_{EW}\varphi|$ pulse pairs without an observer chosen offset value.

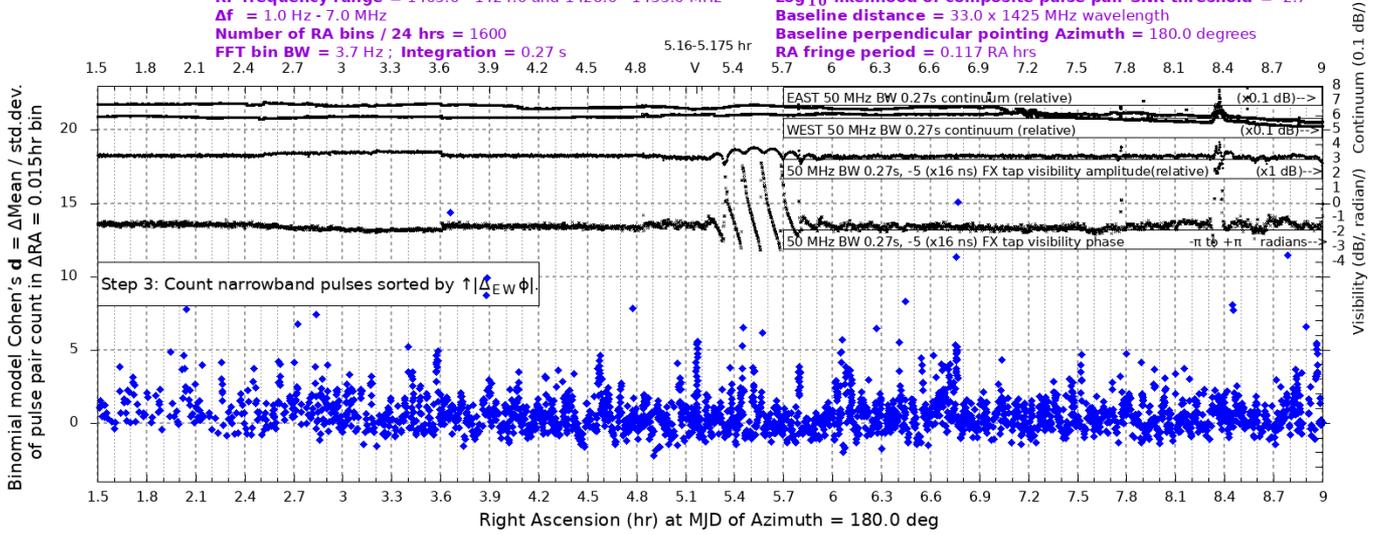

**Figure 3: Step 3:** Sorting the measurements by increasing value of |Δ$_{EW}$φ|, at zero residual phase offset, provides a measurement associated with RA angle relative to the perpendicular bisecting plane of the interferometer baseline. Aliasing due to [-π:π] phase wrapping may result from a single celestial source indicated in multiple RA bins. The statistical power measurement indicates approximately 4 and 5 standard deviations of the mean pulse pair count in two RA directions, conjectured to be one aliased direction, near 5.16-5.175 hr RA. The isolated sporadic outliers in Cohen's **d** are expected due to random near-zero |Δ$_{EW}$ φ| measurements at the top of the sorted pulse heap. These outliers naturally have a high Cohen's **d** due to the low probability of their presence in 3,417 points, while not showing evidence of RA repetition within the sorted heap.

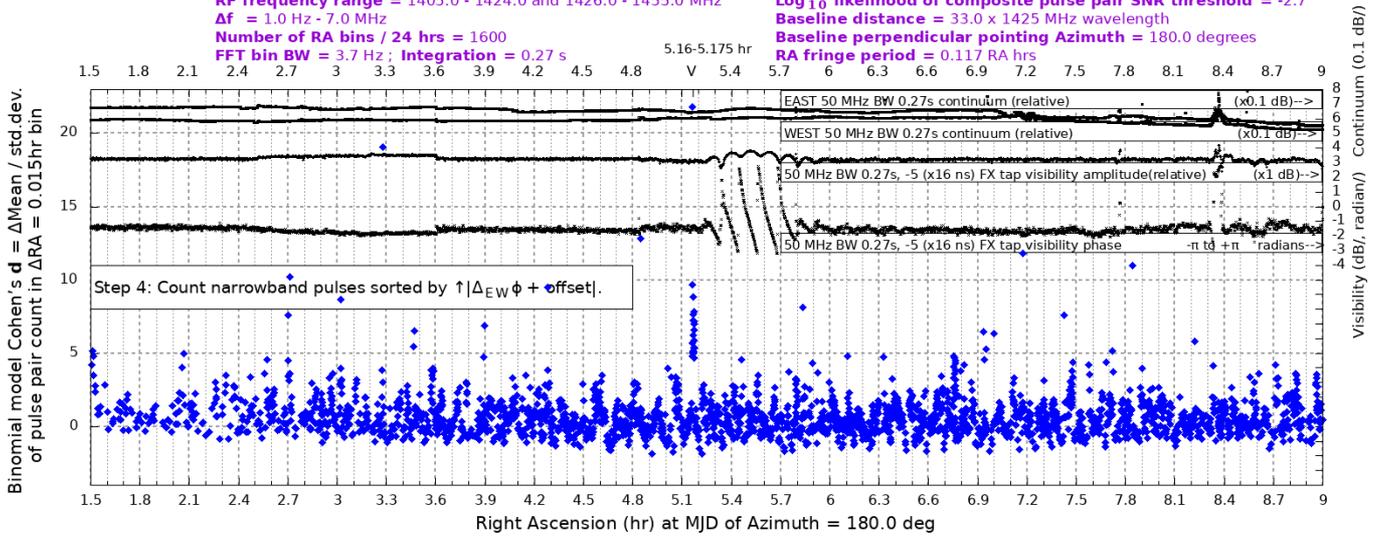

**Figure 4: Step 4:** Sorting the pulse pair measurements by increasing value of |Δ$_{EW}$φ + **offset**|, using the observed residual phase offset apparent in **Fig. 2**, results in a peak Cohen's **d** at 21.8 standard deviations, and twenty values between 4.7 and 9.7 standard deviations. Calibration object Orion A, at 5.5 hr RA, does not appear to be a source of RA-concentrated increases in narrow bandwidth pulse pair counts. **Fig. 8** describes the 5.16-5.175 hr and nearby RA ranges in detail.

A proposed signal discovery method in interstellar communication

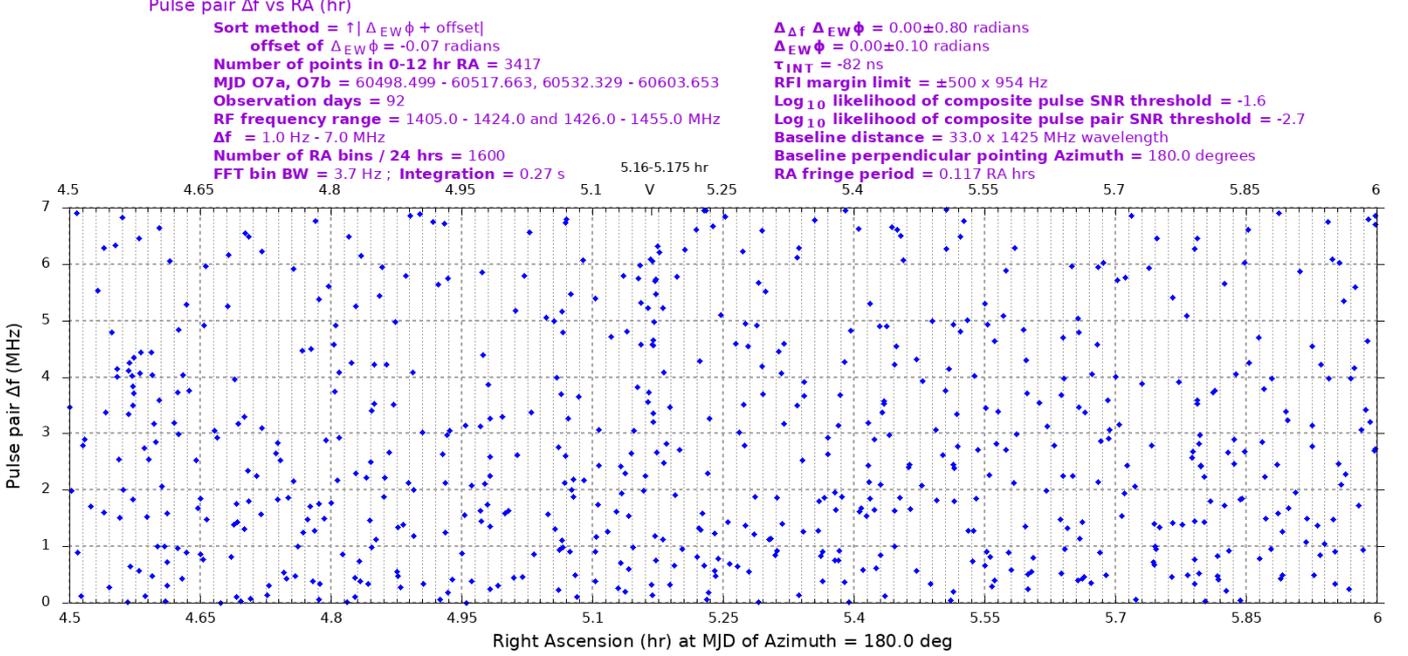

**Figure 5:** Measurement of the frequency separation of simultaneous pulses in pulse pairs appears to indicate a concentration in the RA direction of interest, 5.16-5.175 hr, in the range of **Δf** ≈ 3 to 6.5 MHz. This observation was used, together with an estimate of noise-caused **Δf** values to have a mean count ≈ 5 MHz, to set a post-processing **Δf** filter at 1 Hz to 7 MHz.

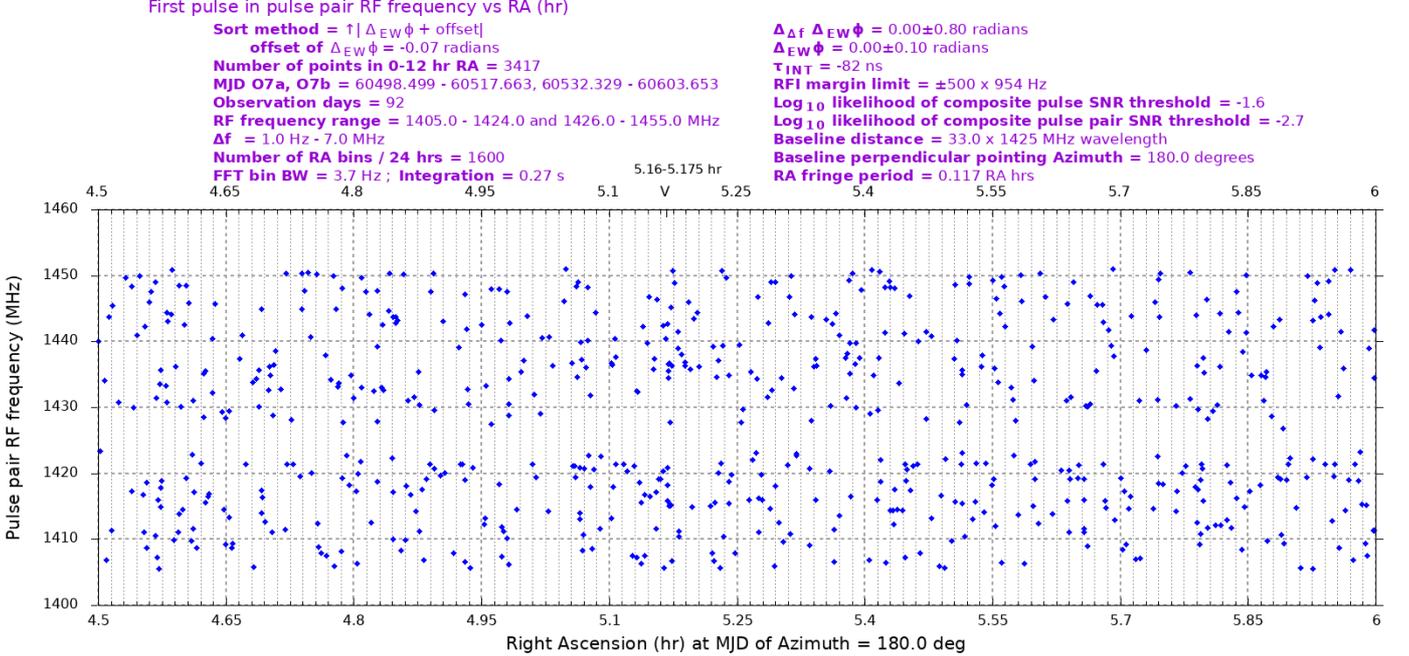

**Figure 6:** Examination of RF frequency of candidate pulses is useful to troubleshoot potential RFI causes. Horizontal lines across RA bins would generally be expected to be apparent if band-limited RFI is present. In an energy-efficient, high information capacity communication system, signals are expected to be distributed across a wide RF frequency spectrum.[1] Distribution of intentionally transmitted pulses having low $\Delta_{EW}\varphi$ are expected to group at RF frequencies that have spacing equal to $1/\tau_{INT}$, where $\tau_{INT}$ is the instrument differential delay. Grouping in RF frequency depends on the $\Delta_{EW}\varphi$ phase noise contributions. The analysis of associated pulse pair measurements and signal models is ongoing work.

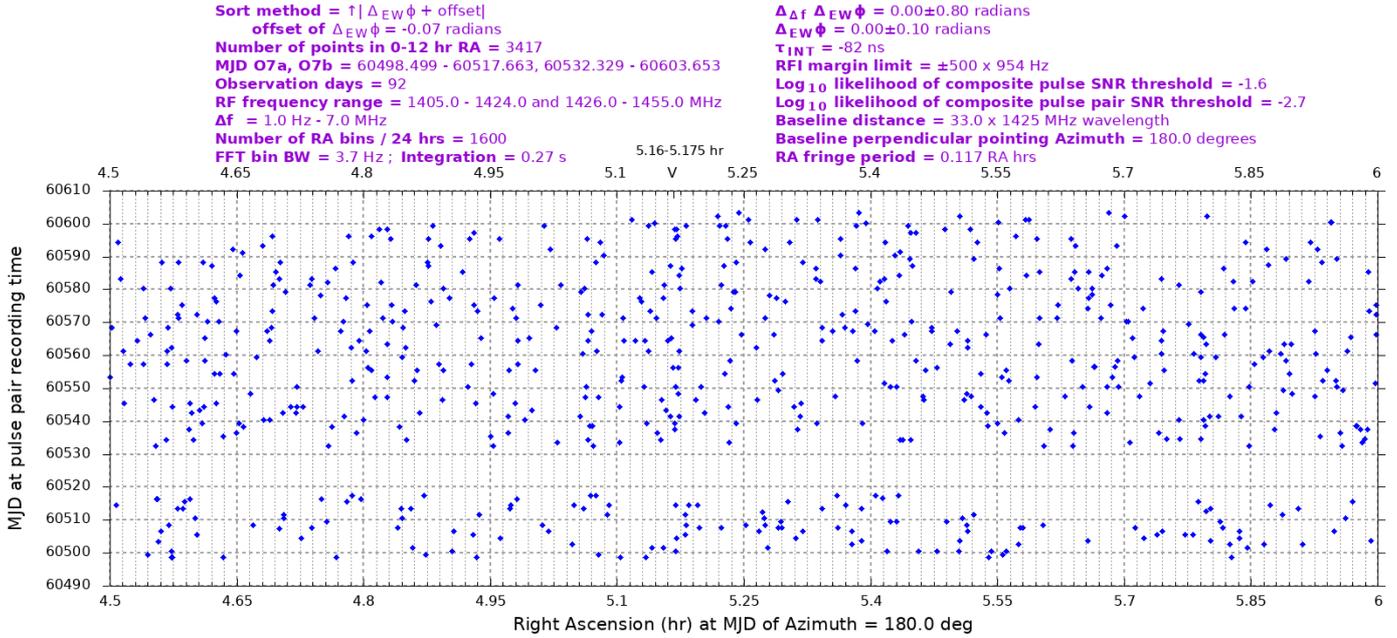

**Figure 7:** A distribution of pulse pairs across MJD days does not indicate an unusual concentration of pulse pairs on the same day. Examination of the 21 data points in the 5.16-5.175 hr RA bin shows that MJD days 60541, 60556 and 60598 each indicated two pulse pairs during the same day. The apparent sparse temporal concentrations of pulse pairs may contribute evidence to an intent of a transmitting entity to avoid interfering, suggested in **III. TRANSMITTER DESIGN, Reduce interstellar transmitter-caused interference. [5]**

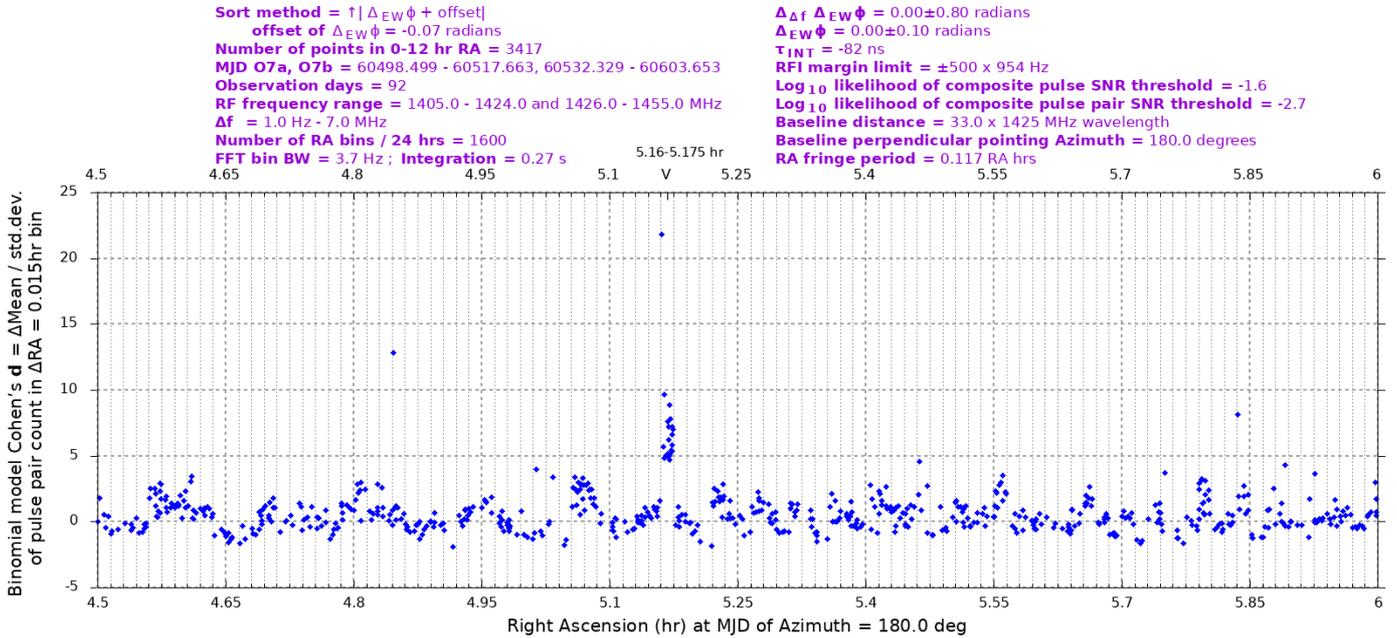

**Figure 8:** A close analysis of the 5.16 to 5.175 RA bin indicates that anomalous pulse pair count is present in this bin throughout the post-processed dataset of 3,417 pulse pairs, spanning 92 observation days. This pulse pair indication may be due to a continuously transmitting, narrow bandwidth pulse source in the 5.16 to 5.175 hr RA direction. Possible RA-aliased clusters appear at -6 and -7 RA bin offsets below the RA bin of interest. 7 RA bin offsets correspond to a 0.105 hr RA aliased period, less than the 0.117 RA value calculated based on instrument calibration. The 21.8 standard deviation value has measurements of 11.79 dB SNR on the East antenna and 11.44 dB SNR on the West antenna, well above the 8.5 dB SNR threshold, and measured 0.0701 radians $\Delta_{EW}\phi$, explaining its large Cohen's **d** value, given the **Fig. 2** derived $\Delta_{EW}\phi$ instrument offset of -0.070 radians.

Additional measurements of the 21.8 standard deviation pulse pair: $\Delta_{\Delta f}\Delta_{EW}\phi$ = +0.41 radians, **MJD** = 60564.439135, **RF frequency** = 1419.207393 MHz, $\Delta f$ = 2.253599 MHz, **RA** = 5.160117 hr. The 21.8 standard deviation pulse pair was the first pulse pair in the 3,417 point dataset, sorted by increasing values of |$\Delta_{EW}\phi$ + **offset**|.

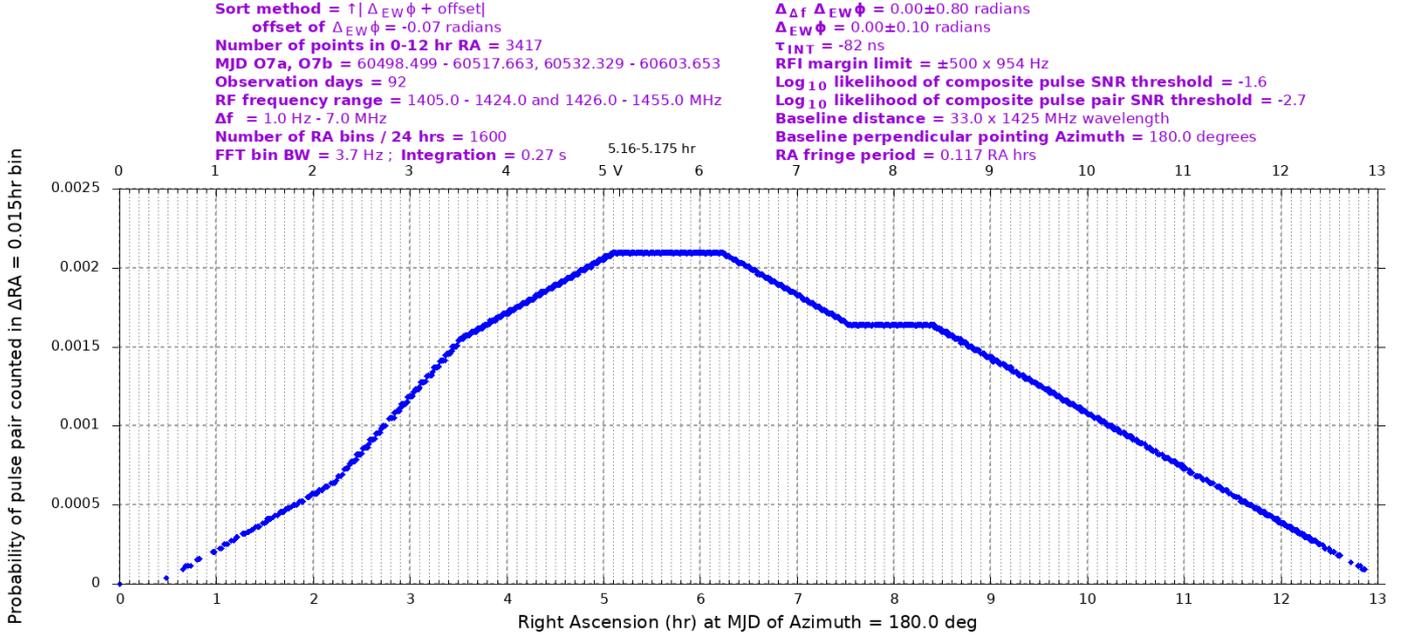

**Figure 9:** The probability of an event counted in each RA bin has a shape affected by the outage period, MJD 60517.663 to 60532.329, when the interferometer telescope system was powered off. The relative RA bin event probability across RA bins does not bias the values of Cohen's **d** across RA bins because post-processing calculates the AWGN model-expected mean count and AWGN model-expected standard deviation, of each RA bin, given each RA bin event probability.

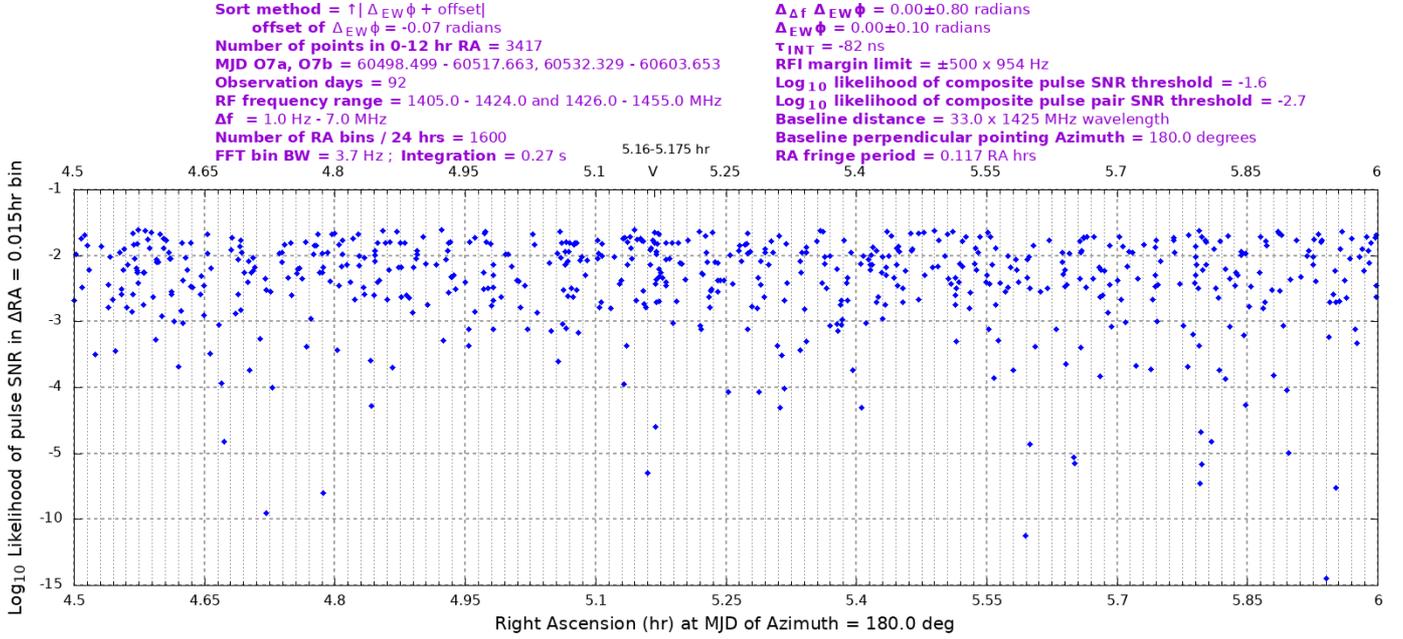

**Figure 10:** Pulse composite SNR and pulse pair pulse pair composite SNR are calculated and converted to a log likelihood representation, **Equ. (2),** to facilitate SNR probability estimation, assuming an AWGN model. Likelihood is based on a Rayleigh-distributed signal amplitude model, having an exponential power probability distribution, while assuming an AWGN cause. Log likelihood is referenced to zero at 8.5 dB SNR. Pulse composite SNR combines the East and West SNR values of the first pulse in the pulse pair. **Fig. 10** displays the pulse log likelihood, while **Fig. 11** displays the pulse pair log likelihood. The log likelihood of SNR distribution affects the observer choice of SNR threshold. SNR threshold setting is an important aspect of receiver design because noise-induced false positives confuse statistical analysis. In addition, high SNR pulses are expected to be correlated with low phase noise, important when measuring repetitive angles of arrival.

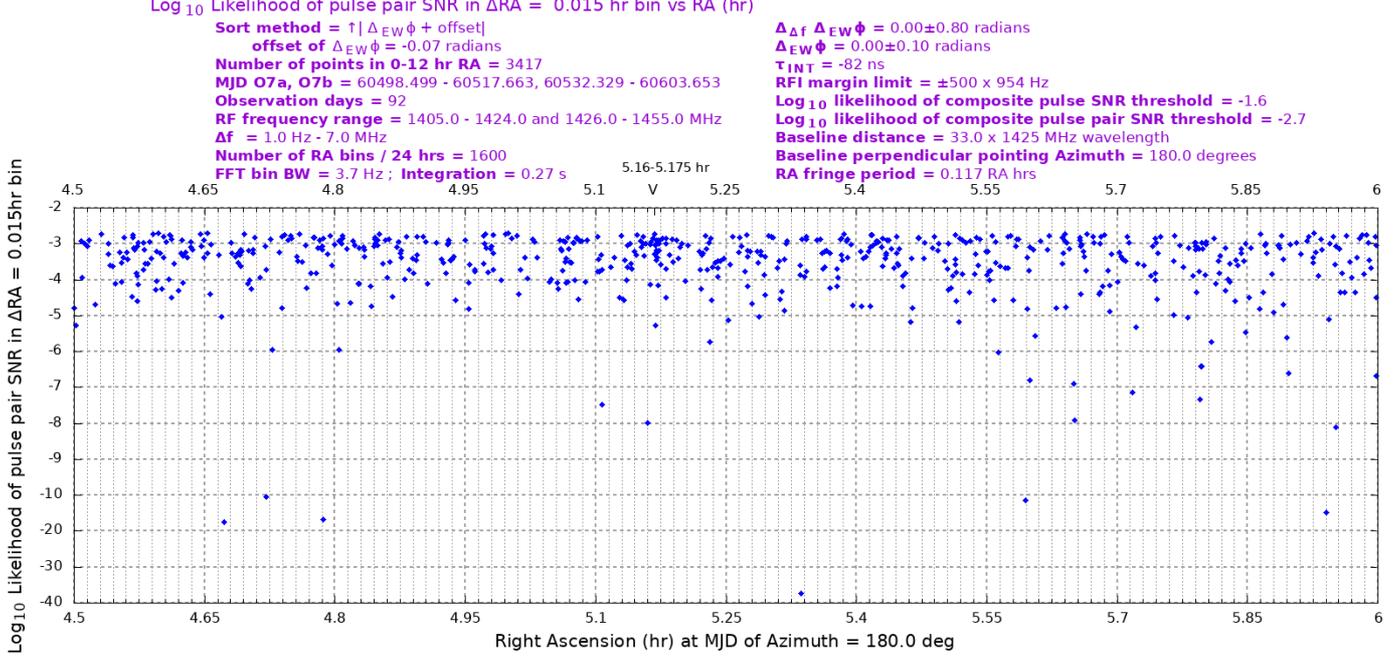

**Figure 11:** Pulse pair log likelihood of composite SNR appears to indicate a concentration of pulse pairs in the RA 5.16 to 5.175 hr direction of interest. Pulse pair composite SNR combines the East and West SNRs of the two **Δf** spaced pulses in the pulse pair, for a total of four SNR measurements contributing to the log likelihood SNR calculation. The pulse pair likelihood filter is set to a more negative value than the pulse log likelihood SNR threshold, to have similar noise-rejection properties.

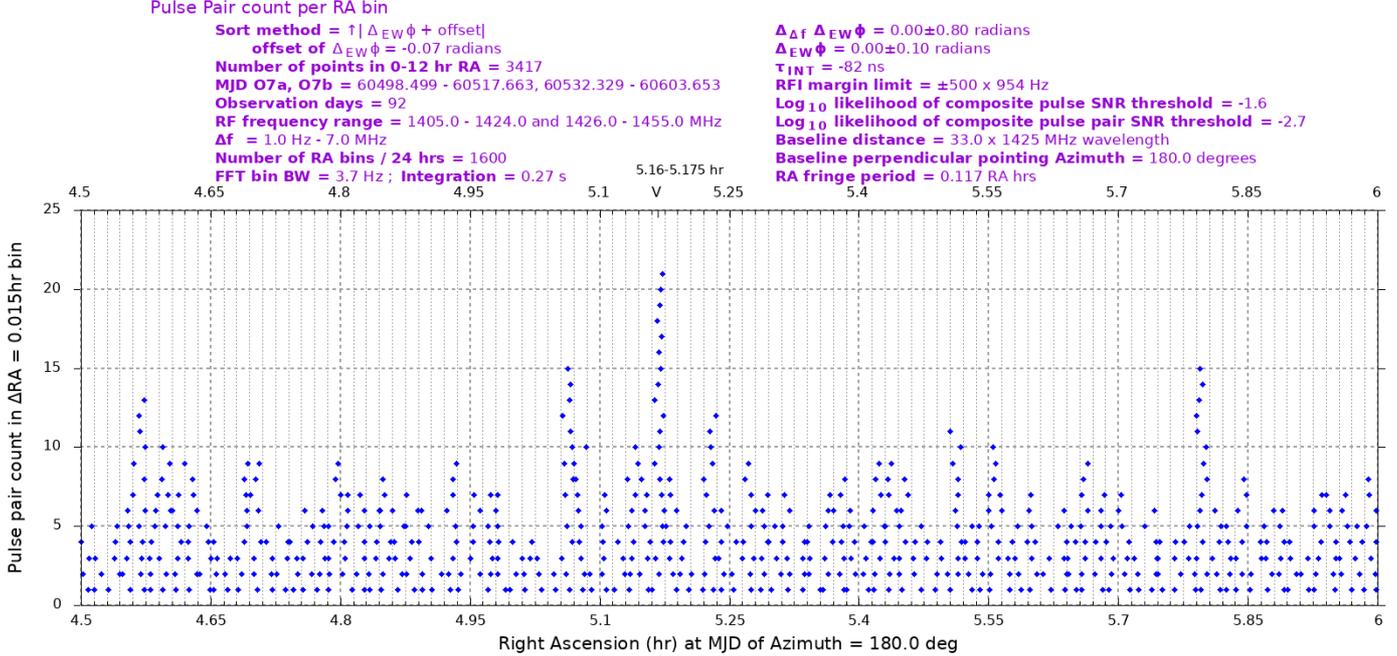

**Figure 12:** Examination of pulse pair count in RA bins indicates that approximately 5 of the 21 pulse pairs in the 5.16 to 5.175 hr RA bin may be attributed to noise. Possible RA aliasing is apparent at 7 RA bins below the 5.16 to 5.175 hr RA bin.

## V. DISCUSSION

### Experimental settings

An objective in this work is to minimize the number of instrument settings that are adjusted during the measurement of Δt=0 Δf polarized pulse pair count over RA. Rather, adjustments are made based on instrument metrology, communication theory and engineering principles.

In this work, the following post-process filter settings are present:

**Δf filter**

The 8.5 dB SNR threshold is used to store Δt=0 Δf pulse pair measurements in the first-level processing files, and results in approximately ten pulse pairs stored per three second interval. Ten pulse pairs in a 50 MHz bandwidth result in an average noise-caused **Δf** estimated at 5 MHz. The **Δf** filter was set to accept pulse pairs having frequency separation from 1 Hz to 7 MHz.

**Instrument delay**

$\tau_{INT}$ was measured at -82 ns, at 1425 MHz, using a test antenna placed on each antenna reflector surface, and measured with a vector network analyzer with instrument-calculated port extensions. Several contributions to delay and phase shift were not included in the measurement, e.g. downconverter delay difference, non-linear frequency dependent delay, transition reflections, and digitizer phase shift across East-West and IQ channels.

The measurement of $\Delta_{EW}\phi$ is affected in two ways by the value and uncertainty of the instrument delay measurement, considering how signal phase may be accumulated, as follows:

$$\frac{\partial \phi}{\partial \omega} = \tau \quad \text{and} \quad \frac{\partial \phi}{\partial \tau} = \omega \, , \qquad (3) \text{ and } (4)$$

where $\phi$, $\omega$, and $\tau$ are instantaneous values of signal phase, angular frequency and delay in a signal propagation path. The first partial derivative indicates phase variation with frequency changes, at a given $\tau_{INT}$ setting. The second partial derivative indicates phase variation with uncertainty in $\tau_{INT}$, while the frequency is constant.

The measured instrument delay, $\tau_{INT}$ = -82 ns, is not corrected in the measurement of $\Delta_{EW}\phi$, due to metrology presently insufficient, given **Equ. (4),** to provide confident low residual system phase offset after instrument delay correction, over 50 MHz bandwidth. Rather, near-zero $\Delta_{EW}\phi$ phase measurements, of wide occupied bandwidth transmitted signals, are expected, due to frequency-aliasing to near-zero phase values, due to $\tau_{INT}$, aliasing across the 50 MHz measured bandwidth, during long duration experiments. Phase noise contributions that cancel each other may also produce near zero phase values.

In the measurement of $\Delta_{\Delta f}\Delta_{EW}\phi$, instrument delay $\tau_{INT}$ = -82 ns, is applied in the measurement, because **Equ. (3)** calculates to less than 0.05 radian phase uncertainty, with 1 ns uncertainty in the measurement of $\tau_{INT}$, while frequency differences within pulse pairs range between 1 Hz to 7 MHz.

The correction and calibration of $\Delta_{EW}\phi$ and $\Delta_{\Delta f}\Delta_{EW}\phi$ measurements, without introducing observer-chosen parameters, are topics of further work.

**SNR thresholds**

SNR thresholds are set above the 8.5 dB first-level processing SNR threshold, because the Ricean distribution contributes to the detectability of high SNR signals. [18] SNR thresholds were adjusted to reduce the number of pulses in statistical analysis, under the assumptions that higher SNR pulses are less likely to be noise-caused and more likely to provide low SNR-induced phase noise.

**Phase measurement filter thresholds**

Chosen $\Delta_{\Delta f}\Delta_{EW}\phi$ and $\Delta_{EW}\phi$ filter thresholds are displayed in the figures, approximately reflecting their phase noise, estimated in **III. Measurements,** $\Delta_{EW}\phi$**,** and $\Delta_{\Delta f}\Delta_{EW}\phi$.

**RFI margin threshold**

Margin in RF frequency to suspected RFI was set to 500 RFI segments, each segment having a 954 Hz span, to reduce possible RFI sideband response.

**RA bin size**

RA bin size has been reduced from 0.1 hr ΔRA in prior work, to 0.015 hr ΔRA, to improve RA measurement resolution.

There is some discrepancy between the O5 and O6 peak Cohen's **d**, observed at 5.3-5.4 hr RA, shown in Fig. 1 and Fig. 2 in **[10]**, and the 5.16-5.175 hr RA bin measured in this report. This may be due to instrument delay-induced phase differences as a result of different length RF cables, between O5, O6, and O7a, O7b observations.

### Hypotheses discussion

**Intentional extraterrestrial transmitter**

The likelihood that the phenomena described in this experiment may be explained by an intentional extraterrestrial transmitter is reduced significantly due to the many decades during which prior searches have not found repetitive narrow bandwidth celestial signals. On the other hand, it seems possible that a few celestial sources transmitting infrequent, short duration, narrow bandwidth pulses, may have been considered to be outliers of Gaussian noise, or RFI, due to not having been subsequently observed.

The increasing human use of secure and secret communication systems suggests that a human-made transmitter might explain the pulses.

The consistent presence of narrow bandwidth pulses, apparently sourced from a single celestial direction, during a 92 day observation, is a countervailing argument, to suggest that the transmitter may be extraterrestrial. As with false positives caused by RFI, pulsars, and fast radio bursts, close examination of the phenomena and theory of source emission is required before one can confidently hypothesize an extraterrestrial communication transmitter. For example, the model of a human-made secure communication transmitter requires explanations of how the transmitter avoids motion in celestial coordinates, due to our solar system's gravity wells.

In general, 0.27 s duration, 3.7 Hz bandwidth directional interstellar signals are expected to be generated by producing a plane electromagnetic wave across a large transmitting aperture, while ensuring that the transmitting aperture has low celestial-coordinate rotation rate, to retain energy in the signal's bandwidth. Channel capacity of such a system may be increased by transmitting pulses in various bandwidths, over a wide bandwidth. Wide-spaced transmitting antennas may be phased to produce multiple non-interfering communication channels to multiple receivers on a planet.

Many receiver experiments are required to increase the confidence in an extraterrestrial transmitter hypothesis.

Hypotheses related to equipment causes and experimental procedures are reasonable, based on numerous ways that false positives might be produced, e.g. software errors, hardware issues, spurious instrument signals, RFI leakage.

**Natural object Doppler spread**

An omni-directional narrow bandwidth radiator, having a rotating spherical radiating surface, and an equatorial circumference of 3.7 wavelengths, will produce a ±3.7 Hz Doppler spread when rotating at one revolution per second. The maximum absolute value of Doppler spread occurs at ±$C_\lambda R$, **[14]** where $C_\lambda$ is the spherical circumference measured in wavelengths, and R is the revolution rate in revolutions per second. As the $C_\lambda R$ product increases, a 3.7 Hz bandwidth receiver becomes less sensitive to the coherent energy that the sphere is producing. Natural object models need to include means to avoid Doppler spread desensitization.

**Independent verification**

There is a possibility that verifications of the hypothesis proposed in this work might falsely indicate replication of results, if the verification methods use a problematic aspect of the experimental method. It therefore seems important that verification be designed and conducted as independently as reasonably possible.

**RA comparison to observations in 2018 and 2019**

Two synchronized radio telescopes, having forty foot diameter in Green Bank, West Virginia and having sixty foot diameter near Haswell, Colorado were used to seek simultaneous reception of near-equal Doppler-corrected RF frequencies of 3.7 Hz bandwidth, 0.27 s duration pulses at -7.6 degrees DEC.**[5]** FWHM antenna beam widths are 1.0 and 0.6 degrees respectively.

Two of three, and two of seven, of the highest SNR values recorded, during two 48 hour observation runs, measured RA at 5.184, 5.253, and 5.156, 5.271 hr RA, respectively, among a 24 hr range of RA observation. The Green Bank and Haswell pointing directions calculate sky angles within -0.06 and 1.45 degrees of those of the limits of the 5.16 to 5.175 hr RA bin observed and reported here.

## VI. CONCLUSIONS

The hypothesis of this experiment, i.e. a noise cause, has been falsified to a high statistical power level, leading to further work to examine and test auxiliary and alternate explanations relevant to the narrow bandwidth pulsed signal phenomena.

An interstellar communication explanation of the measured phenomena has not been falsified by experimental evidence.

Independently developed replication is required to aid in the development of explanatory hypotheses.

## VII. FURTHER WORK

1. Independently developed measurements are required to provide confidence that the results reported here correctly represent hypothetical celestial transmitted signals.

2. Associated measurements and aliased responses are expected to provide ideas for the development of signal source models and hypotheses. Natural object model development is required.

3. Analyze the phase noise contributions and probability distributions of phase measurements to aid in the settings of filter thresholds.

4. Post process directions outside the 1.5 to 9 hr RA range reported here, to seek similar phenomena.

5. Improve instrument metrology.

6. Repeat experiments, using the same and different receiver settings.

## VIII. ACKNOWLEDGEMENTS

During the long-term work on this project, many workers in many organizations have contributed ideas and encouragement, including Green Bank Observatory, National Radio Astronomy Observatory, Society of Amateur Radio Astronomers, SETI Institute, Deep Space Exploration Society, Allen Telescope Array, The Penn State Extraterrestrial Intelligence Center, Breakthrough Listen, Berkeley SETI Research Center, The SETI League, HamSCI, the open-source community and equipment vendors. Special thanks also go to my friends and family, and to my co-workers at communication and test equipment manufacturers, for their generous help and encouragement.